\documentclass[12pt]{article}

\usepackage{graphicx}
\usepackage[tight,TABTOPCAP]{subfigure}

\begin{document}

\title{Study of the resistivity in a channel with dephased ripples }

\author{Gabriel Arroyo-Correa$^{1}$, Ivan Herrera-Gonzalez$^{2}$, \\
 Alberto Mendoza-Su\'arez$^{1}$, 
Eduardo S. Tututi$^{1} \footnote{ e-mail:tututi@umich.mx}$\\
{\it $^{1}$Facultad  de Ciencias  F\'{\i}sico-Matem\'aticas}, \\
{\it Universidad Michoacana}, \\ 
{\it 58060 Morelia, Mich., M\'exico}\\
 {\it $^{2}$Instituto de F\'{\i}sica y Matem\'aticas}, \\  
{\it Universidad Michoacana,}, \\ 
{\it 58060 Morelia, Mich., M\'exico}}

\maketitle


\begin{abstract}
We study  the transport properties of classical particles 
in the ballistic regime trapped in a two-dimensional channel with dephased
ripple boundaries. By taking into account small ripple amplitudes
an analytical  approximate expression for the classical resistivity
is obtained. We show that the resistivity can be increased considerably
by dephasing the  walls of the channel. Our results are compared with those
obtained for a channel composed of a flat  and a sinusoidal boundaries.
\end{abstract}



\section{Introduction}

The development of  nanotechnology, the physics of  thin films along with the 
theory of dynamical systems have  motivated the study of transport properties 
of billiards \cite{alhassid,nandini,meyerovich,marcus}. Two-dimensional billiards 
consisting on a  point  particle moving freely in a two-dimensional region bounded by rigid walls. The billiards problem  leads to  both classical and quantum  Hamiltonian systems such as the well known stadium \cite{stadium} and Sinai  billiards \cite{sinai}. 
A  question of great interest in  open billiards, such as  those composed 
of quantum dots and  quantum wires, is that concerned with  the influence of the boundary shape on the transport properties of this systems 
\cite{berry,burki,nakamura,kouwenhoven}.

The study  of chaos by means of  classical transport properties is
of much   interest recently. So,  some classical transport properties   
have been  analyzed in open channels with  different boundaries \cite{sanders,baowen,casati,alonso-vega,jepps}. For example a criterion was proposed
 to distinguish between regular and chaotic dynamics by measuring the classical resistance 
in a two-dimensional  channel composed of  two boundaries being one sinusoidal 
(rippled) and the other a flat. We shall be referring to as a semiplane channel  \cite{luna1}.  The quantum  counterpart  of this system was 
studied in Ref. \cite{luna2} where the energy band spectra, the eigenfunctions and 
the quantum Poincar\'e sections were obtained for a free particle moving inside the 
channel.  It was proposed 
in Ref. \cite{luna4} the construction of a microlaser of highly directional emission 
by using a two-dimensional  semiconductor  waveguide that is composed of  two 
semi-infinite leads connected by a cavity.

Some preliminary experimental results concerned with semiplane  channel has 
been published recently \cite{luna5}. It has been found that in order to get 
a high reflectivity  is necessary to have a big number of trapped particles

In this paper we analyze  the effects of the 
boundaries on the properties  of classical ballistic transport for  a 
two-dimensional channel composed of  two  sinusoidal boundary walls as shown in Fig.  
\ref{figu1} (a). Our main objective in this  work is to analyze  how the relative 
phase and amplitude of the ripple of the walls  affect the
transport properties of the channel and the possible applications to  waveguides. 
An interesting feature of our system is that it has a  high reflectivity,  
compared  with the obtained using the semiplane channel.

A brief outline of this paper is as follows. In section \ref{channel}
we describe the geometry of the channel and give the corresponding map. In section \ref{poinca} we present a study of dynamics of the system  by means of  the Poincar\'e sections. Section \ref{properties} is devoted to obtain  analytical expressions 
for the resistivity when  amplitudes of the ripples are small. 
In section \ref{numerical} we present numerical results for the reflectivity and transmitivity. Finally in section \ref{concluding} we give the concluding remarks.

\section{The  channel and the map}\label{channel}

\begin{figure}
 \centering
\subfigure[]{\includegraphics[scale=0.3]{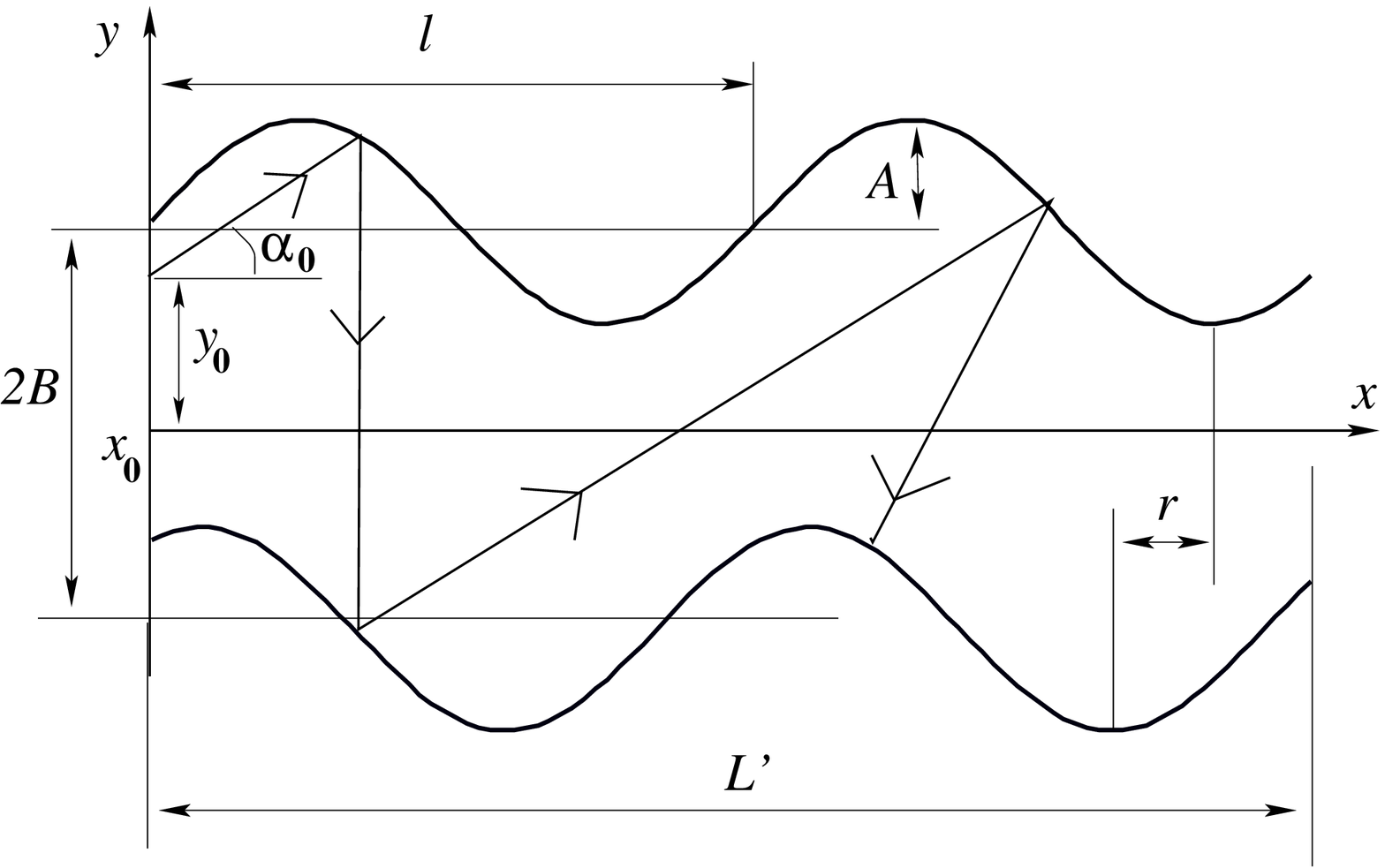}}\qquad
\subfigure[]{\includegraphics[scale=0.25]{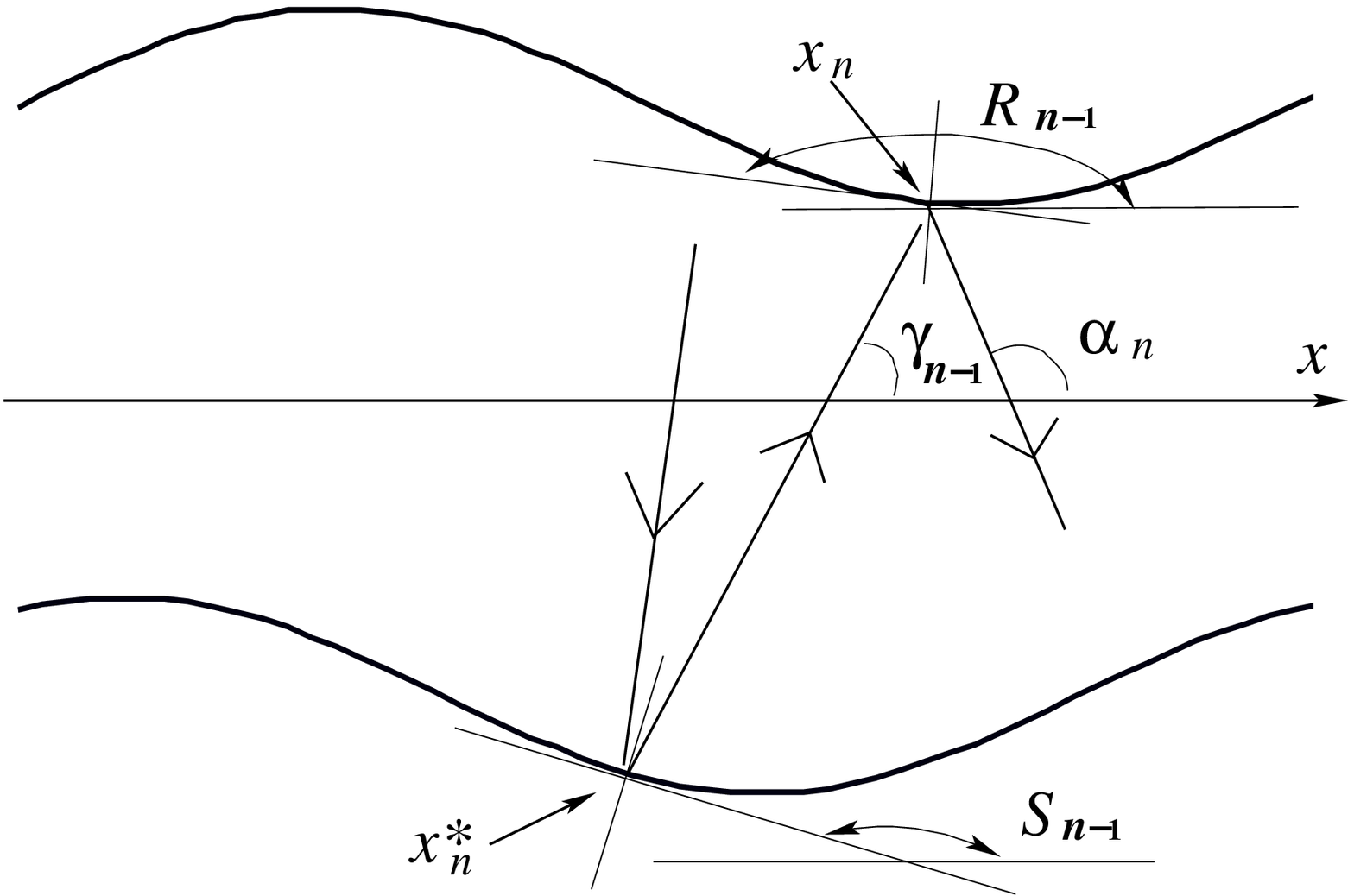}}
\caption{(a) The channel. (b) The parameters used in the map.}
\label{figu1}
 \end{figure}

Let us consider an open channel with rippled walls (see
Fig. \ref{figu1} (a)). The profiles of the  upper and lower walls are determined 
respectively by 
\begin{eqnarray}
y_1 & = & b + a \sin 2 \pi x
\nonumber\\
y_2 & = & -b +a\sin 2\pi\left( x+r\right), 
\label{walls}
\end{eqnarray} 
where we use the dimensionless variables 
$x=\frac{X}{l}$, $y=\frac{Y}{l}$,  $b=\frac{B}{l}$, $a=\frac{A}{l}$, 
$L={L^{\prime}\over l}$,
with $l$ being the length of one period,  $L^{\prime}$ is  the length of the 
channel, $A$ is the amplitude of the ripple and $B$ is a half of  the average 
width of the channel.  The variable $r$ denotes the  phase difference between 
the upper  and lower walls.   As a consequence of  the periodicity of boundaries of 
the  channel,  $r$ is restricted to  values $0\leq r<1$. 

We are interested in determining the evolution of $n_0$ particles injected on the 
left side of this channel. The particles are dropped by $N$ sources uniformly distributed
along the $y$ axis located at $x=x_0$. Each source drops $\frac{n_0}{N}$ particles with   
an angular distribution   given by
\begin{equation}
\rho(\alpha_0)=\frac{n_0}{2D(x_0)}\cos(\alpha_0),
\label{distri1}
\end{equation}
where $D(x)=y_1(x)-y_2(x)$ is the distance between the walls at the  $x$-point.
 We assume that the collisions of the particles with the boundaries are  specular.

\subsection{The map}
In order to find the Poincar\'e sections corresponding  to channels characterized 
by certain parameters  it is necessary  to have a  discrete map. 
Since the parameters we shall be using in the next section are such that
multiple collisions of the particles with the ripple  are  improbable.
According to our numerical study   the occurrence of multiple collisions,
for each trajectory  is $ \sim 10^{-3}$,  therefore we may  neglect them.

To construct the map let us consider a set of discrete points 
$(x_n,\alpha_n)$ of each particle  initially dropped with  initial conditions 
$(x_0,y_0,\alpha_0) $. Here $x_n$ is the position in the  $x$ direction at the $n$-th 
collision of the particle with the upper wall and $\alpha_n$ is the angle that the trajectory
of  particle  makes with $x$-axis  just after the $n$-th collision. 
The resulting map is
\begin{eqnarray}
&&\alpha_{n+1}=2\left(S_n-R_n\right)+\alpha_n
\nonumber\\
&&(x_{n+1}-x^{*}_n)\tan\gamma_n +a\sin 2\pi( x^*_n+r)=2 b+a\sin 2\pi x_{n+1}.
\label{map}
\end{eqnarray}
From the second line of last equation we  see that  to express the position of the
$(n+1)$-th collision as  $x_{n+1}=f(x_{n}, \alpha_n)$, the  equation must be 
solved numerically. The variables $\gamma_n$, $R_n$, $S_n$ and $x^*_n$ are given 
 by (see Fig. \ref{figu1} (b)):
\begin{eqnarray}
&& \gamma_n =\pi-\alpha_n + 2S_n
\nonumber\\
&& R_n =\tan^{-1}(\frac{dy_1}{dx})\vert_{x_n}
\nonumber\\
&& S_n =\tan^{-1}(\frac{dy_2}{dx})\vert_{x^*_n}
\nonumber\\
&& 2b+(x^*_n-x_n)\tan\alpha_n + a\sin 2\pi x_n = a\sin 2\pi( x^*_n+r).
\label{angles}
\end{eqnarray}
Here $x^*_n$ stands for position of the particle in the $x$-direction just  
after the $n$-th collision with the 
lower wall, $\gamma_n$ is the angle that the trajectory of the particle  makes with 
$x$-axis  after the $n$-th collision with  the lower wall.
Finally, $\tan S_n$ ($\tan R_n$) represents  the slope of the line tangent to the lower  
wall at the point $x=x^*_n$ (the upper wall at $x=x_n$).

We want to indicate for future comparison,   that our channel (for  $r=\frac{1}{2}$) has
 the same  transmitivity as the  semiplane channel  with an average 
width equals to $b$ and the same ripple amplitude. In order to see this, let us consider a particle  propagating inside the semiplane channel and other one propagating inside our channel. The trajectory of the particle
moving in the semiplane channel defines a succession  of collision points ($x^p_n,y^p_n$) with the upper wall. On the other hand,  a particle propagating in our  channel, with   $r=\frac{1}{2}$  defines a set  of collision points with the upper wall ($x_n,y_n$). Because of the specular  symmetry respect to the $x$ axis in our channel, the following condition is  
satisfied 
\begin{displaymath}
 (x^p_{2n+1},y^p_{2n+1})=(x_n,y_n), \qquad n=1,2\dots,
\end{displaymath}

if  the particles in  both channels are dropped with the same initial conditions and 
the first bounce occurs with the upper wall (if the first bounce is with the lower wall 
a similar relation is obtained) then it is obtained the same transmitivity in both cases.

\section{Poincar\'e sections}
\label{poinca}
To plot  the Poincaré\'e sections  we  choose for convenience 
the conjugate pair ($x_n$,$p_n$), where $p_n=\cos \alpha_n$ 
(see the map in  Eq. (\ref{map})) is the  momentum in the $x$ direction right 
after the $n$-th collision with the upper wall. In order to have access to all possible orbits
with our initial conditions
we varied the position of the sources at the $x$ coordinate. For instance, to reach orbits within the islands we placed the sources at the $x$ coordinate of the corresponding fixed point. Notice that we implicitly take
a channel of infinite length.

For obtaining the Poincar\'e  plots  we set the value $l=1$ and 
accordingly we have a  narrow channel
with $b=0.1$  and a wide  channel  with $b=2.5$. To represent the
Poincar\'e plots, we use the $x$-interval $[0,2]$.  We will see that 
the dynamics of these two channels are  quite  different. To solve the map in Eq. (\ref{map})
we use a bisection method applied to $10^5$ small intervals along the channel. 
The numerical code was written in Fortran and the calculations were carried out in a PC with
a  Pentium IV processor.

\begin{figure}
 \centering
\subfigure[]{\includegraphics[scale=0.3]{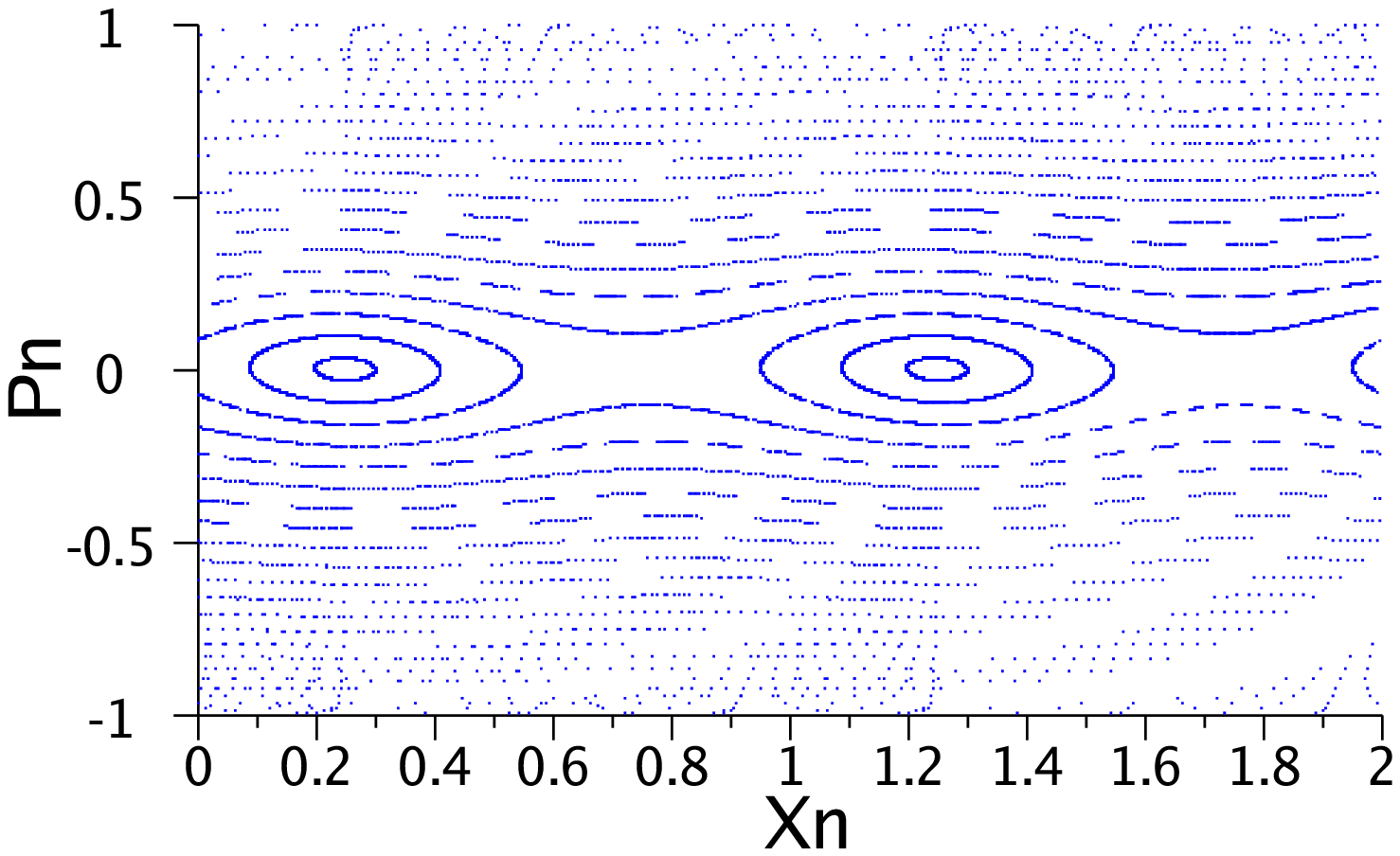}}\qquad
\subfigure[]{\includegraphics[scale=0.3]{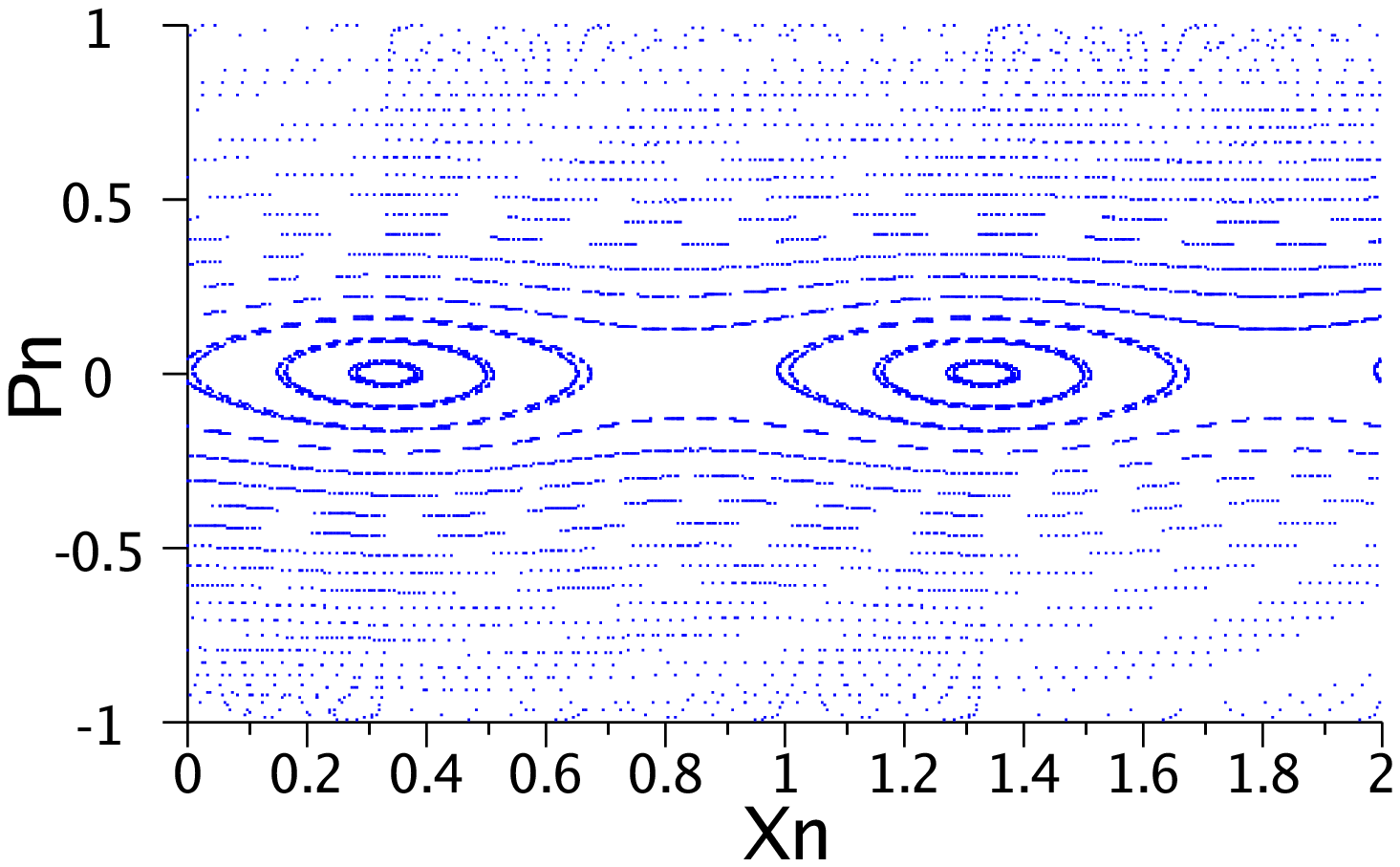}}\\
\subfigure[]{\includegraphics[scale=0.3]{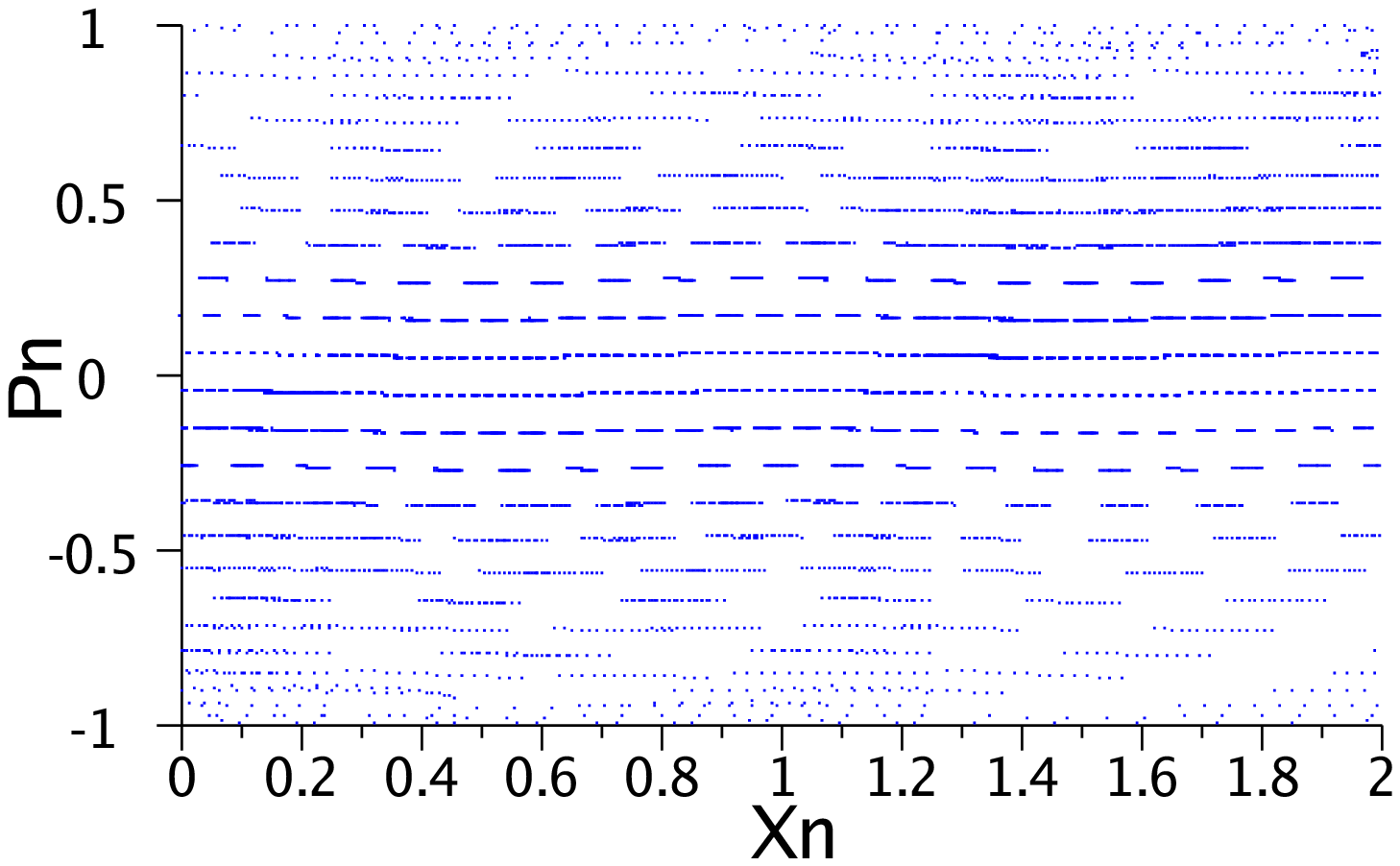}}\qquad
\subfigure[]{\includegraphics[scale=0.3]{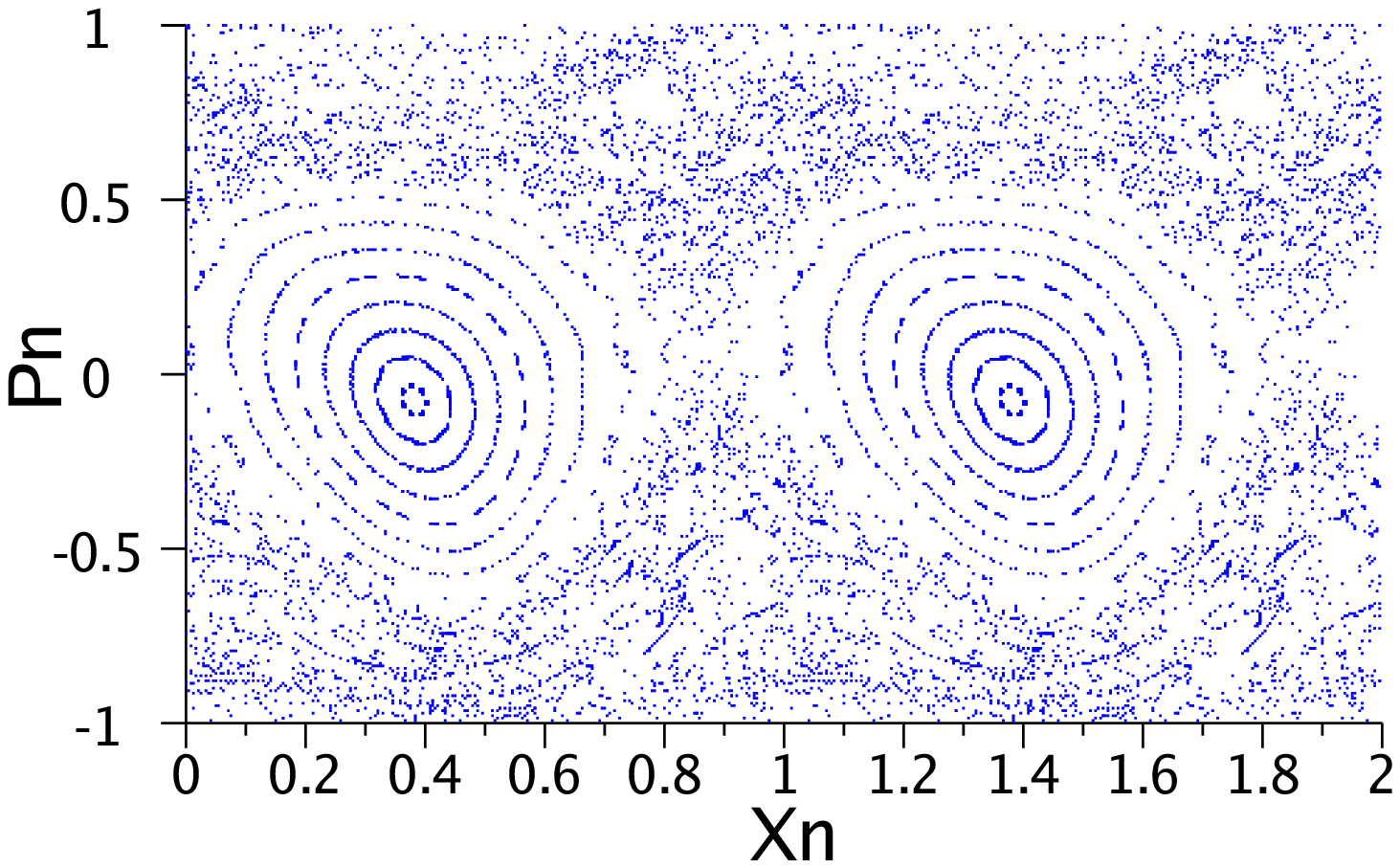}}\\
\caption{Poincar\'e plots for the narrow channel.  (a) $a=0.001$ and $r=\frac{1}{2}$, 
(b)  $a=0.001$ and $r=\frac{1}{3}$, (c) $a=0.001$ and $r=0$ and (d) $a=0.017$ and 
$r=\frac{1}{4}$.}
\label{lnc1}
 \end{figure}

 For the narrow channel with  small ripple amplitudes   (e.g $a=0.001$), the  
 system shows a regular dynamics;  the corresponding Poincar\'e plots resemble  
the phase space of a one dimensional pendulum  (see Figs. \ref{lnc1}(a) and (b)). 
This means that the elliptic orbits  (an elliptic orbit is  an special
case of a closed trajectory in the phase space   called  librational motion \cite{goldstein}) correspond to  trapped  particles 
in the channel,  moving backward and forward around a stable fixed point. 
The trajectories  outside these elliptic orbits represent particles traveling  to the 
left  ($p_n<0$) or to the right ($p_n>0$) of the channel and they never return. 
We can also see that the position of the fixed points and the size of the region of librational motion  depend on the  relative phase $r$.  For  instance, with $r=0$ the  elliptic orbits cannot be observed (see Fig. \ref{lnc1} (c)).  For other cases the size 
of the elliptic orbits are clearly remarked. In our case fixed points  represent particles bouncing between the  same point ($x_f$,$p_f$) of the upper wall and  the point ($x^*_f$,$p^*_f$) of the lower wall. The position of fixed points $(x_f,p_f)$ can be obtained by considering 
geometrical arguments (see \ref{appen}).

\begin{figure}
 \centering
\subfigure[]{\includegraphics[scale=0.3]{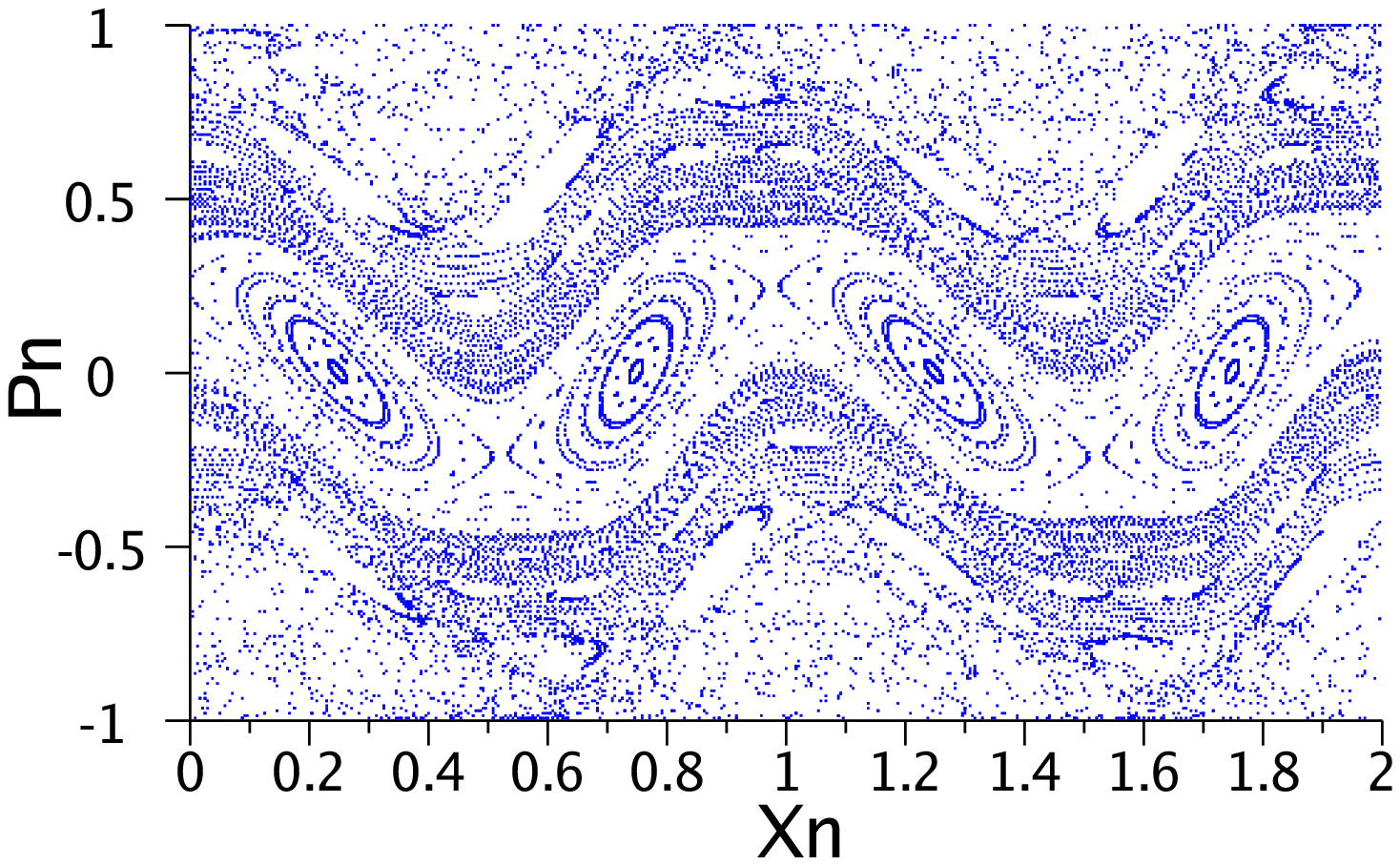}}\qquad
\subfigure[]{\includegraphics[scale=0.3]{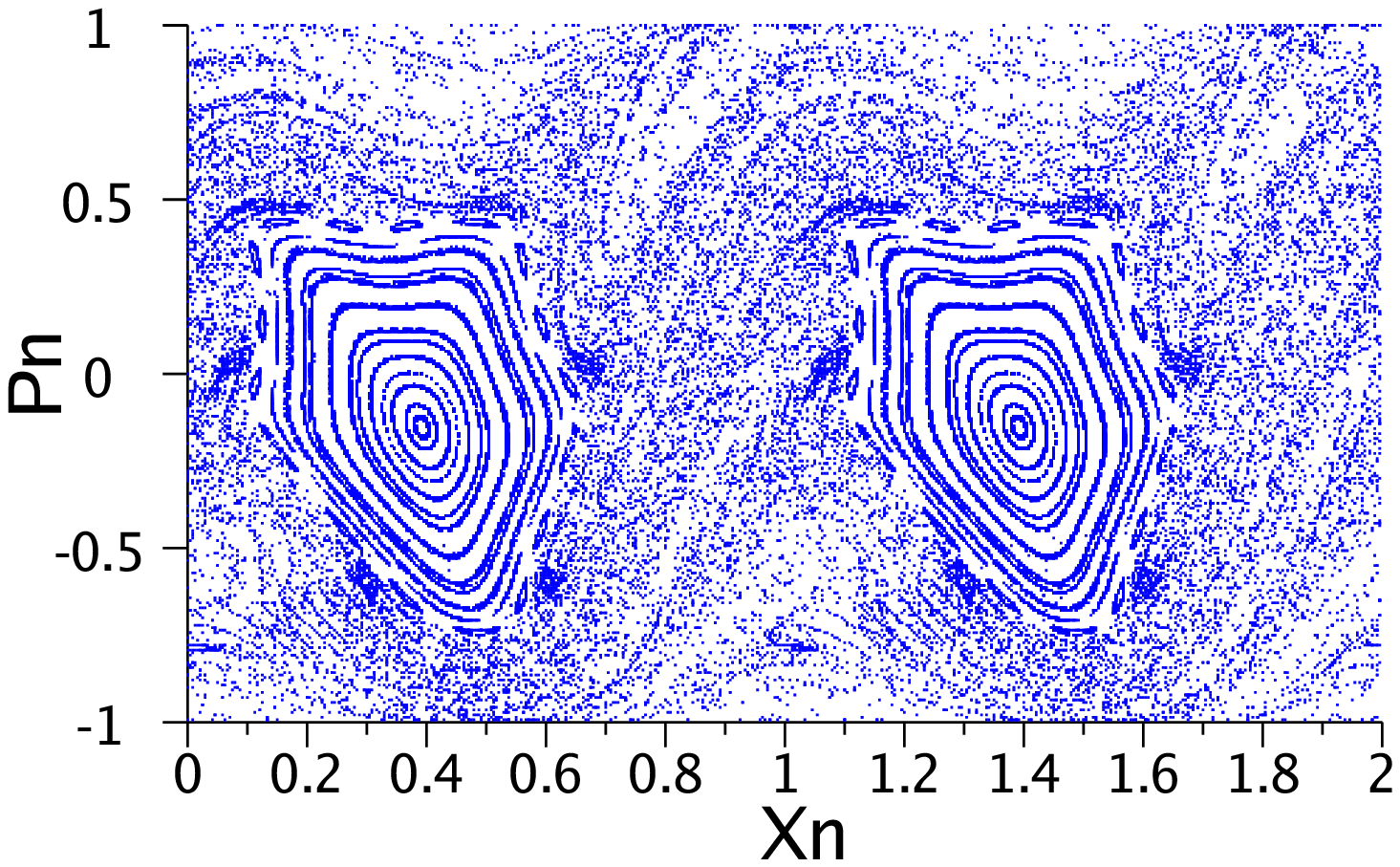}}\\
\subfigure[]{\includegraphics[scale=0.3]{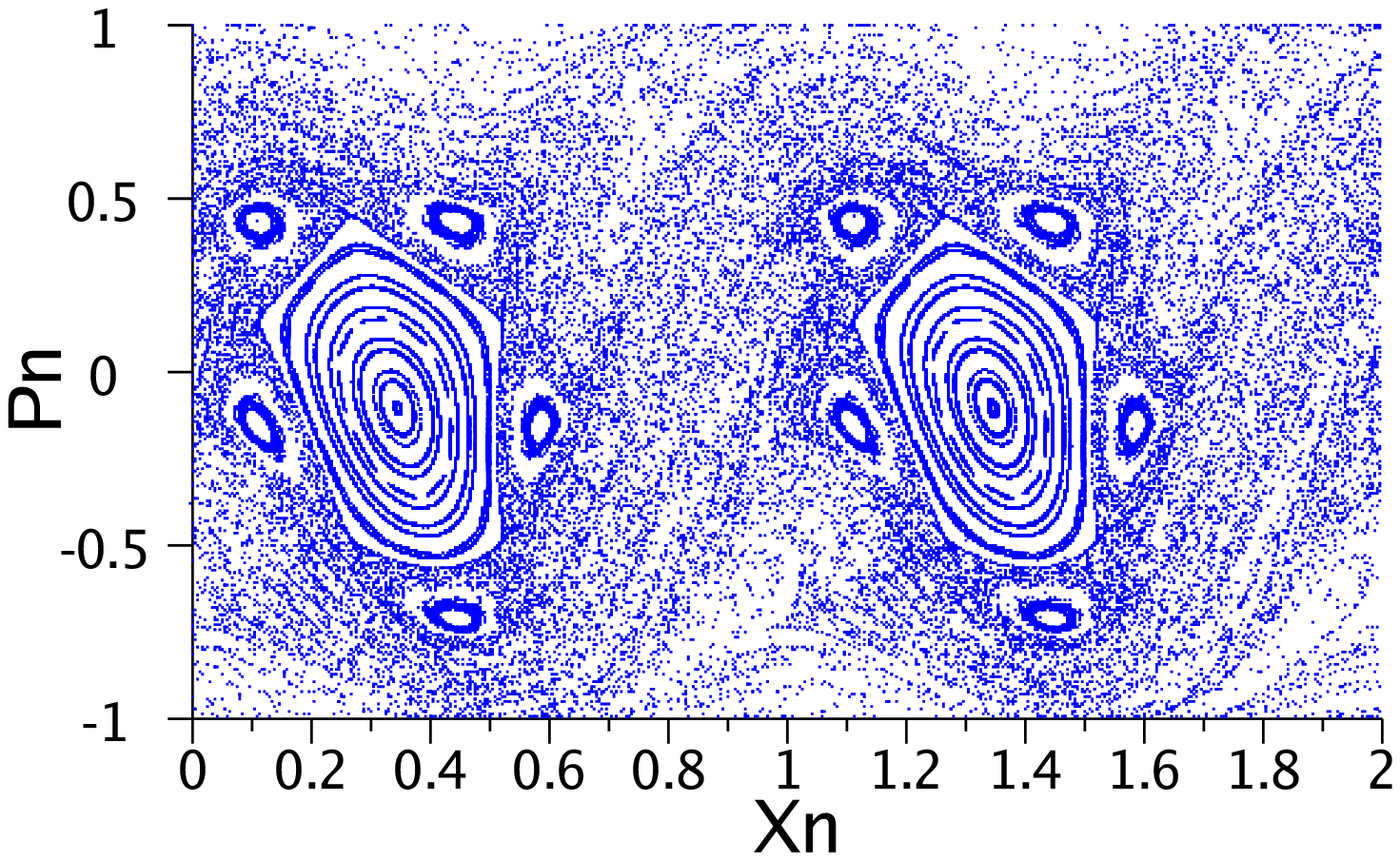}}\qquad
\subfigure[]{\includegraphics[scale=0.3]{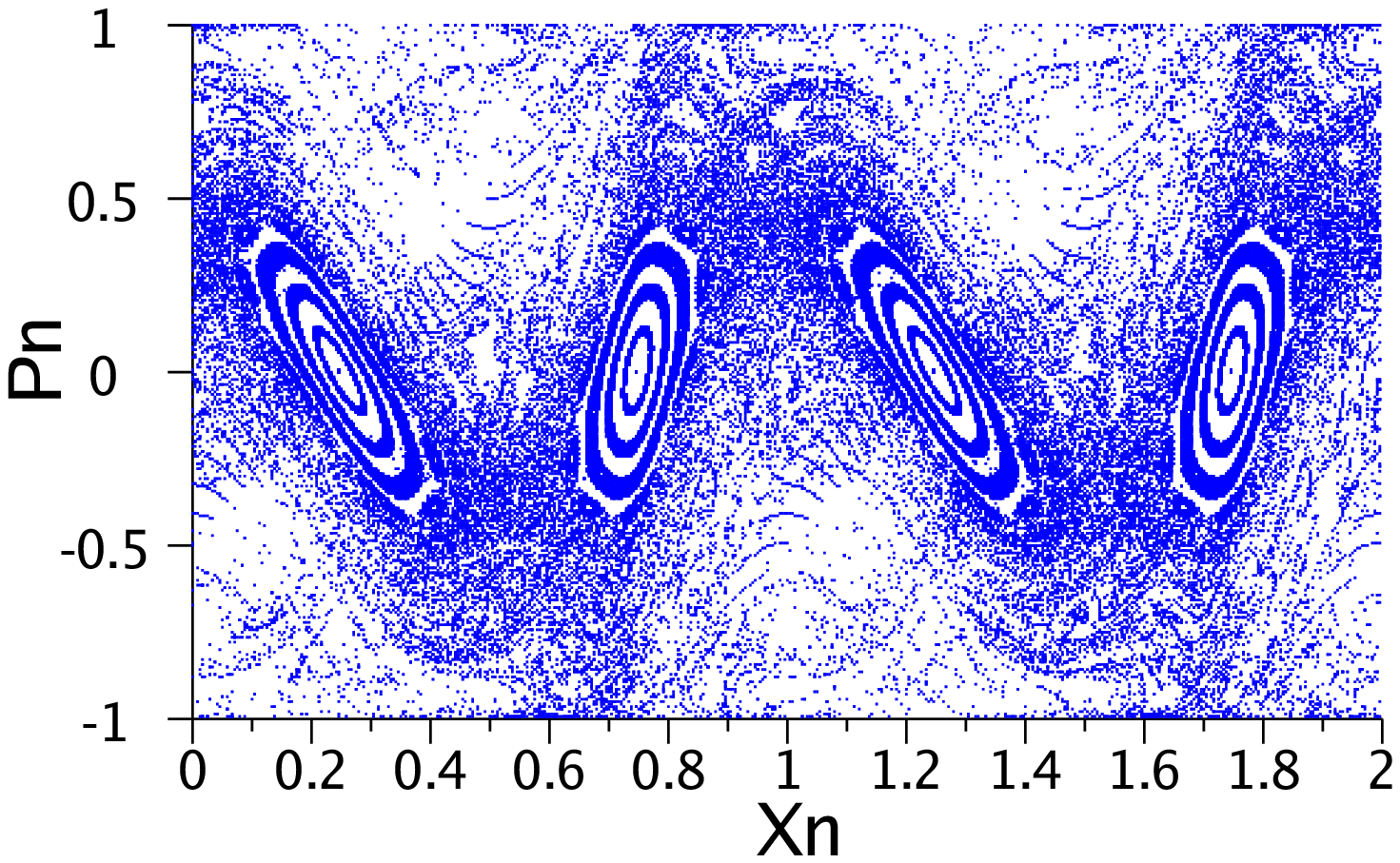}}\\
\caption{Poincar\'e plots for the narrow channel.  (a) $a=0.04$ and $r=0$, 
(b)  $a=0.032$ and $r=\frac{1}{4}$, (c) $a=0.03$ and $r=\frac{1}{3}$, and (d) $a=0.065$ 
and $r=0$.}
\label{lnc2}
\end{figure}

If we  increment $a$ (still for a  narrow channel) in some interval (e.g. for the values $r=0.5$, $0<a< 0.009$),  the 
dynamics of the system is still regular  and the arising librational  orbits occupy a larger region. This means that there are more  librational orbits accessible to the initial conditions increasing the number of  reflected particles in the channel.  In the case of $r=0$ and $a=0.01$, the elliptical orbits can now be  observed. If we increase $a$ 
again, the elliptical orbits start to deform,
occupying a larger  region and the separatrix becomes chaotic with some sizable width 
(see Fig. \ref{lnc1}(d)), this means that more particles can get into the region and 
they may be reflected. The accessible chaotic region outside  the separatrix can not 
contribute to the reflection because there are some KAM curves that forbid its connection with the separatrix (see Figs. \ref{lnc2} (a) and \ref{lnc1} (d))
\footnote{Notice that the lower  wall can in principle reflect particles. 
That is the reason why we can have elliptic orbits  around a stable fixed point 
of  period one  without crossing the $x$ axis 
(see Fig. \ref{lnc2} (b)). In the case of the  semiplane channel the 
trajectory of the particle in the phase space must cross the  $x$ axis to have some  reflection.   Hence,  it could be thought 
that there can be reflection of particles without crossing the KAM curves, but this does 
not occur because the  Poincar\'e plots formed with the lower wall as a Poicar\'e plane 
have similar barriers.}. Nevertheless,   there is a critical 
amplitude $a_c$ that depends on  $r$, for which these KAM 
curves break allowing  the connection of all chaotic regions and then they can 
contribute to the reflection (see Figs \ref{lnc2} (b), (c) 
and (d)). For instance   for $r=\frac{1}{4}$,  $a_c=0.015$,  for $r=\frac{1}{2}$, $a_c=0.027$ 
and for $r=0$,  $a_c=0.065$. We  observe  a principal first-order resonant 
island surrounded by  a chaotic
sea, and also  a second order  resonant islands chain surrounding the first order 
islands.  In Fig. \ref{lnc2} (c)  these islands are produced by particles 
bouncing in the neighborhood of a stable fixed point of period five.
If we increase the value of $a_c$, more  initial conditions get into the chaotic 
sea due to destruction of  regular curves.  
Notice that the rate at which the KAM curves are destroyed 
as the amplitude is increased depends on $r$: if we are near to $r=0$ this process 
is slow and if we  are near to $r=\frac{1}{3}$ this process is fast.
%
%

\begin{figure}
 \centering
\subfigure[]{\includegraphics[scale=0.3]{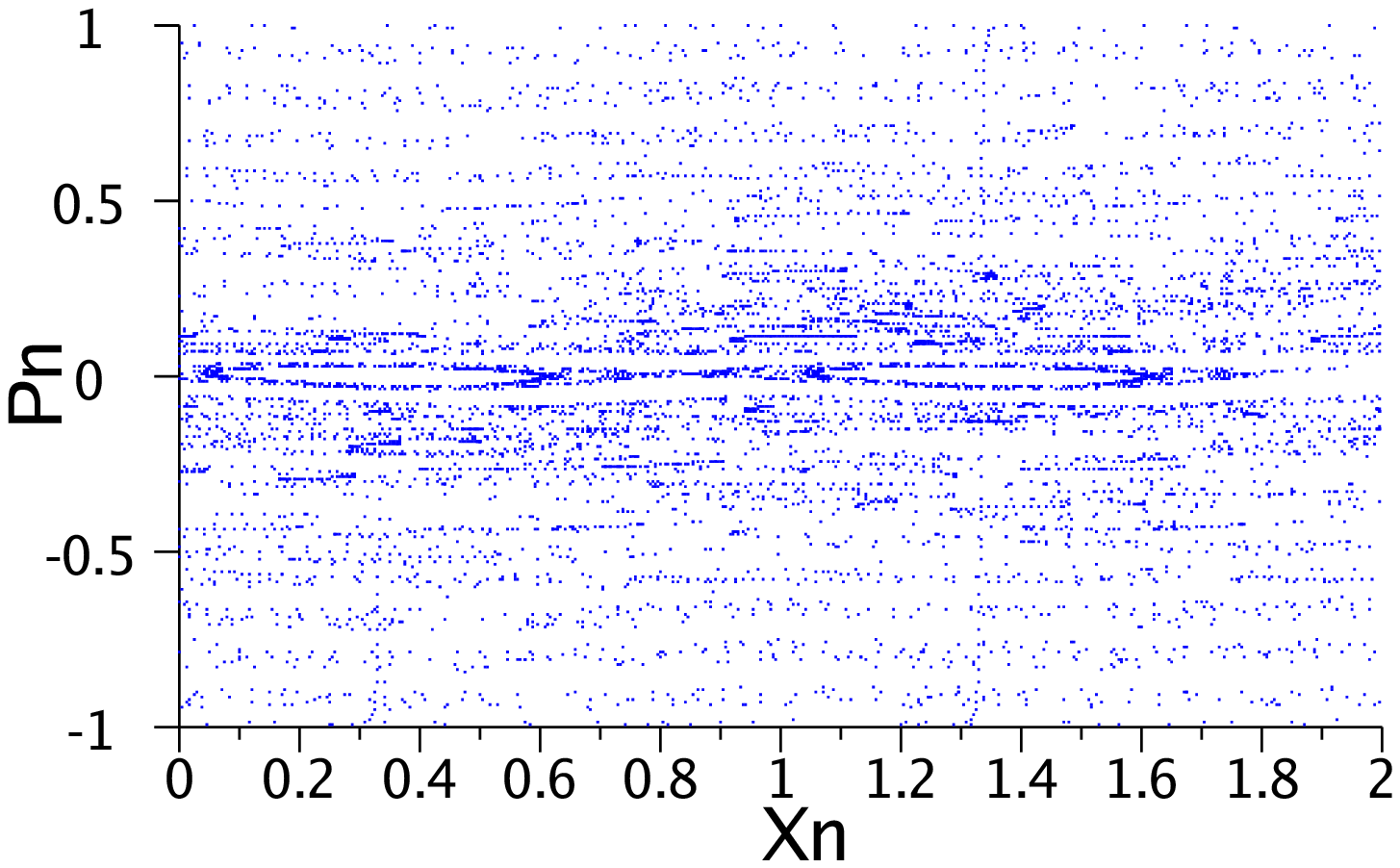}}\qquad
\subfigure[]{\includegraphics[scale=0.3]{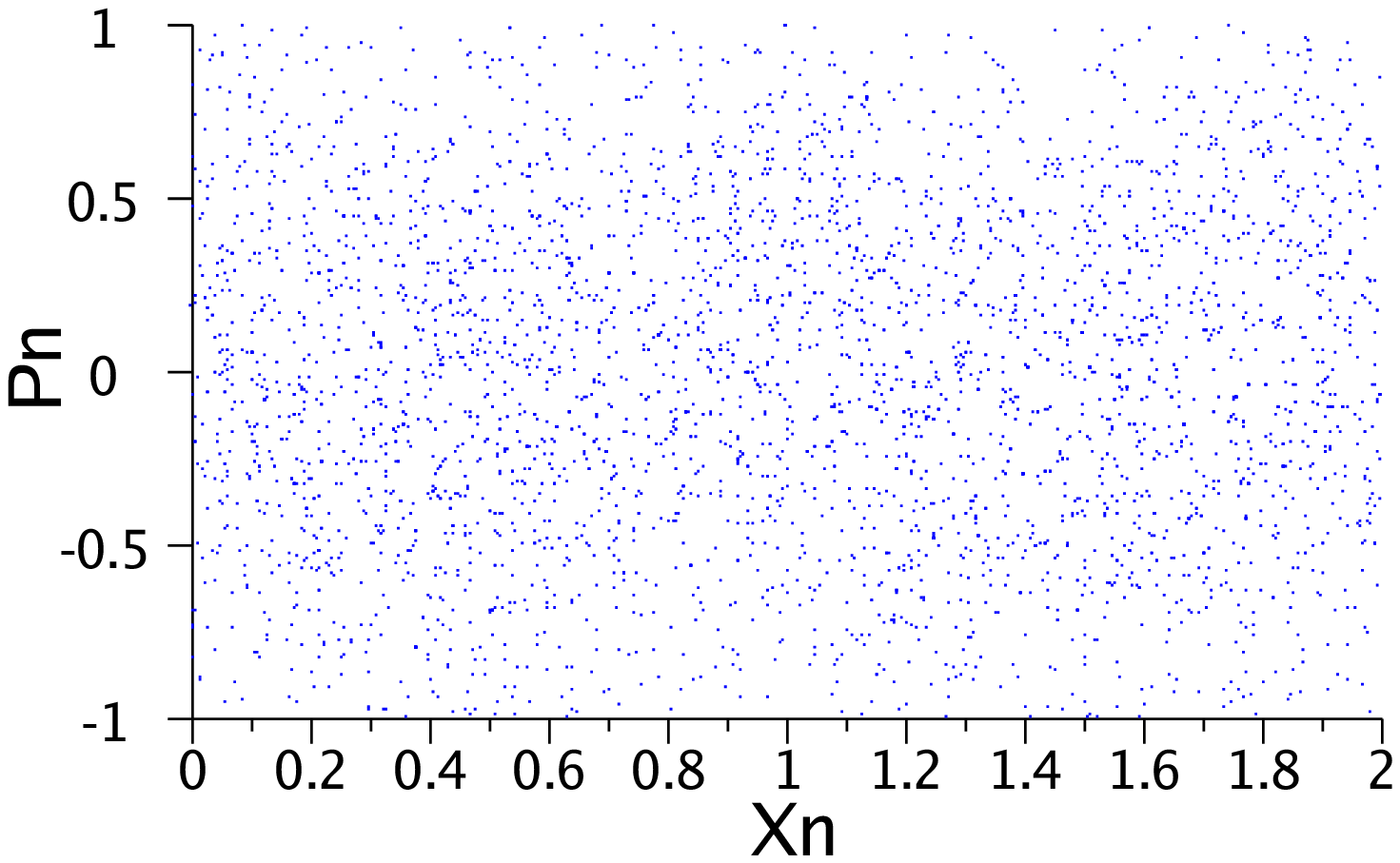}}\\
\caption{Poincar\'e plots for the wide channel.  (a) $a=0.001$ and $r=\frac{1}{3}$, 
(b)  $a=0.008$ and $r=\frac{1}{3}$.}
\label{wnc1}
\end{figure}

In the case of the wide channel, we can see, in contrast to narrow channel, that Poincar\'e plots for the  
wide channel show a rather chaotic behaviour, even for small amplitudes 
(see Fig. \ref{wnc1} (a)), but there are some small regions of librational motion. If the amplitude is increased 
there exists a critical amplitude  for which all KAM curves are 
destroyed and it produces a global  chaos (see Fig \ref{wnc1} (b)). In previous works  \cite{luna4,capasso,bicos} 
it has also been observed that the case of global chaos is generally  meaningless for applications in waveguides. Due to this behaviour
we leave aside the discussion  about this type of channel.

\section{Analytical results for the resistivity}\label{properties}

The transmitivity, $T$, is the flux of transmitted particles divided by the incoming 
flux, as usual the reflectivity is  $R=1-T$. For our system  it is not possible to have analytical expressions for these quantities. However  in some 
limiting cases we can determine them. In what follows we shall be using a finite
channel of length $L=2$.

If we consider  a narrow channel ($b\ll L$) with  small amplitudes 
($a\ll b$) and assume that the ripples are smooth  ($a\ll 1$), it is possible to obtain
an analytical expression   for  reflectivity. In  this  channel,
the particles dropped near to the $y$-direction could contribute
to  reflectivity. In order to obtain an expression for the reflectivity,
let us first consider  two consecutive collisions, for which the particles
collide almost perpendicularly with the walls. Under this assumption we have the
 following
\begin{equation}
 \Delta J \equiv |\sin\alpha_n|D(x_n) -|\sin\gamma_{n-1}|D(x_{n-1}^*)\approx 0.
\label{deltaJ}
\end{equation} 
In fact, according to Fig.\ref{figu1} (b) and Eq. (\ref{angles}) 
the relation  $\alpha_{n}+\gamma_{n-1}=\pi+2R_n $ is fulfilled.
For   angles such that  $\gamma_{n}=\pi/2-\varepsilon$, with  $|\varepsilon |\ll 1$,
considering the $\Delta x=x_n-x_{n-1}^*$, taking into account an expansion
in  Taylor series and keeping terms up to second order in (\ref{deltaJ}) we obtain
\begin{equation}
  \Delta J \approx \frac{dD(x)}{dx}\bigg |_{x_{n-1}^*}\Delta x-2D(x_{n-1}^*)R_n
\left( R_n+\varepsilon\right).
\label{deltaJ1}
\end{equation} 
This equation allows us to corroborate  Eq. (\ref{deltaJ}). Notice that 
$\Delta J=0$ is satisfied trivially for the  case of plane walls. Taking into account
the conditions of the channel in question, we are free to choose the 
parameters in  such a way that  each term in  Eq. (\ref{deltaJ1}) is too small to be 
considered. If we now consider the case between two different collisions,
for instance the  $n$-th and  $m$-th collisions we have that
\begin{equation}
 |\sin\alpha_n|D(x_n)\approx |\sin\gamma_{m}|D(x_{m}^*),
\label{deltaJ2}
\end{equation} 
 under the
same conditions as  previously assumed. To see this, notice that the contribution to reflection comes only from particles whose initial
conditions reach accessible elliptic orbits, colliding almost perpendicularly
with the walls. For this particles the variables  $\Delta x$, $\epsilon$ and $R_n$
in Eq. (\ref{deltaJ1}) can change of sign, in addition, for our conditions
the number of collisions is small, and $\Delta J$ is still small.

By using  Eqs. (\ref{walls}) and (\ref{deltaJ2}), we obtain 
\begin{eqnarray}
&&(2b +a\sin2\pi x_n -\sin2\pi(x_n+r))\cos \beta_n
\nonumber \\
&&=(2b +a\sin2\pi x_m -\sin2\pi(x_m+r))\cos \beta_m,\qquad n\neq m=1,2\ldots
\label{eq4}
\end{eqnarray}
where  $\beta_n=\frac{\pi}{2}-\alpha_n$ is the angle between the velocity and the vertical direction at the $n$-th collision. If we consider particles executing 
librational motion, $\beta_n$  decreases gradually as the particles move
forward until they  reach a turning point, $x_N$, where $\beta_N\approx 0$ 
and then the particle returns. There are two  critical angles $\beta_{c1}$ and $\beta_{c2}$, 
for which the particle does not return.  These  angles correspond to 
particles dropped with  positive and negative angles,  $\alpha_{c1}$ and $-\alpha_{c2}$,
respectively (see Fig.\ref{figu1} (a)). The critical angles  depend on both the geometrical 
parameters and  the initial position ($x_0$,$y_0$) at which the particles are
dropped. For the parameters involved in the narrow channel, we may assume that  $\beta_{c1}\approx\beta_{c2}\equiv\beta_c$, and this condition is  independent
 of $y_0$ (which is corroborated    numerically). To find $\beta_c$ we set the  point
 of departure at $(x_0,y_0) =(x_0, y_1(x_0))$, where $(x_0, \sin \beta_c)$ corresponds to 
an initial point in the phase space that belongs to the largest amplitude of 
librational motion around at some elliptic fixed point $x=x^e_{f0}$. 
Setting $\beta_n=\beta_c$, $\beta_m=\beta_N$, then $x_n=x_0$ and $x_m=x_N\approx x^h_{f0}$ 
is close to an hyperbolic fixed point, whose position is determined by 
$x_0$ and Eq. (\ref{fijopeque}). If we  substitute these values in Eq. (\ref{eq4}) and
expand in a Taylor series by dropping terms of order $ \left( \frac{a}{b}\right)^2 $
and keeping terms of order $\beta^2_c  $, we obtain 
\begin{eqnarray}
(\beta_c)^2 & \approx & \frac{a}{b}\bigg[\sin(2\pi(x^h_{f0}+r))+\sin(2\pi x_0)
-\sin(2\pi x^h_{f0}) 
\nonumber\\
&-& \sin(2\pi(x_0+r))\bigg],
\label{aprx}
\end{eqnarray}    
where $x^h_{f0}$ is a parameter to be determined.
Setting the values $r=\frac{1}{2}$, $x_0=\frac{1}{4}$ in the former equation, then 
$x^h_{f0}\approx \frac{3}{4}$ and  obtain the value for 
$(\beta_c)^2\approx \frac{4a}{b}$. This is the same result previously obtained 
for the   semiplane channel   \cite{luna1}, but  the average width of our 
channel is twice as large.  To find out the general behaviour of $x^h_{f0}$ we use the effective potential introduced below.

According to Eq. (\ref{deltaJ2}),  we have   $C=D(x)|\dot y|\bigg|_{x_n}$ with unit speed 
and $C$ being a constant. Using this condition and the conservation of the energy, 
it is easy to see that the motion in the $x$-direction  is
 described  by \cite{percival}:
\begin{eqnarray}
{\frac{1}{2}}(\dot{x})^2=E-V(x),
\label{eq3}
\end{eqnarray}   
with $E$ being the energy of the particle (with unit mass) and   
$V(x)=\frac{1}{2}|\dot{y}|^2$.
Hence $V(x_n)=\left( {\frac{C}{D(x)}}\right)^2\bigg|_{x_n}$ can be interpreted as an effective potential at  $x_n$ that depends only on the geometric parameters of the channel. The usefulness of this effective potential relies on the fact that it explains
the regular motion and allows one to find out the $x$-position and  the stability of the fixed points. 

 If we use the 
profiles described by Eq. (\ref{walls}), then 
$D(x)=2b+a\sin 2\pi x -a\sin 2\pi\left(x+r\right)$ and the potential  takes the 
form of an oscillatory periodic function. This explains why Poincar\'e plots resemble  
the phase space of a simple pendulum. An interesting case is for $r=0$,  where $V(x)$ is  
constant and consequently, within the approximation used, there are not trapped particles.
This explains  the behaviour of almost straight lines trajectories shown 
in Fig. \ref{lnc1} (c). This situation occurs approximately in  a flat channel.
We can find the $x$ position of the fixed point by obtaining the maxima and 
minima of $V(x)$ which represent the elliptic fixed points $x_f^e$ and 
the hyperbolic fixed points $x_f^h$ respectively.  We obtain that the fixed points  
satisfy the condition:
\begin{eqnarray}
\cos 2\pi x_f=\cos2\pi (x_f +r).
\label{fijopeque}
\end{eqnarray}   
 It is easy to see, for instance, that for $r=1/2$, 
the elliptic and hyperbolic fixed points  are located respectively in 
$x^e_f=\frac{1}{4}\pm n$ and $x^h_f=\frac{3}{4}\pm n$, with $n$ an integer; and that for the case of 
$r=\frac{1}{4}$, the elliptic and hyperbolic fixed points  are located respectively at $x^e_f=\frac{3}{8}\pm n$ and $x^h_f=\frac{7}{8}\pm n$. Let us mention that an elliptic fixed point and the consecutive hyperbolic point are separated by a distance $d_s=1/2$ except for 
$r=0$ where $d_s=1/4$. These  results are in agreement with the Poincar\'e plots obtained in the previous  section.

\begin{figure}[ht]
\begin{center}
\includegraphics[width=6cm]{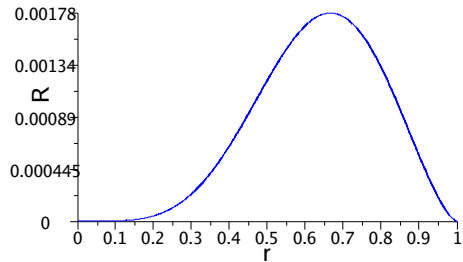}
\caption{Reflectivity of the narrow channel  as a function of $r$, for the value $a=0.001$ and $x_0=0$.} 
\label{resist}
\end{center}
\end{figure}

Taking into account the previous results, we see that the transmitivity takes the form:
\begin{eqnarray}
T =\frac{\int_{-\alpha_c}^{\alpha_c} \rho(\alpha_0)\cos\alpha_0 d\alpha_0}
{\int_{-\frac{\pi}{2}}^{\frac{\pi}{2}}\rho(\alpha_0)\cos\alpha_0 d\alpha_0} 
=1-\frac{2\int_{0}^{\beta_c}\,\sin \beta_0\rho(\frac{\pi}{2}-\beta_0)d\beta_0}
{\int_{-\frac{\pi}{2}}^{\frac{\pi}{2}}\rho(\alpha_0)\cos\alpha_0 d\alpha_0},
\label{eq9a}
\end{eqnarray}     
where $\alpha_0=\frac{\pi}{2}-\beta_0$. Using Eq. (\ref{aprx}) and assuming
$\beta_0$ small, we obtain  
 \begin{eqnarray}
R \approx \frac{4}{3\pi}\beta_c^3= G(r)a^{3/2},
\label{eq10a}
\end{eqnarray}
where  $G(r)$ is given by 
\begin{eqnarray}
G(r)&=&\frac{4}{3\pi \,b^{3/2}}\left[\sin (2\pi(x^h_{f}+r))+\sin(2\pi x_0)\right. 
\nonumber\\
 &&- \left. \sin (2\pi x^h_{f})-\sin(2\pi(x_0+r))\right]^{3/2}.
\end{eqnarray} 
In the former $x^h_f=x^h_f(r)$ is determined from the positions of the sources at $x_0$ and from Eq. (\ref{fijopeque}). Let us emphasize that the length $L$ of our channel  is bigger than the distance
between two consecutive hyperbolic fixed points, this makes the function $G$ 
independent of $L$. In fact, particles executing librational motion cannot escape from the channel
at the right side.  Otherwise, in general,  $G$ is $L$-dependent. In Fig. \ref{resist} we show a graph of reflectivity given by Eq. (\ref{eq10a}). The important point here
is that we can choose the parameter $r$  to obtain a required reflectivity, for instance
for  $r=0.67$ we have that the reflectivity is maximal.

The resistivity $\rho$  of the channel is an important measurable quantity that 
is related to reflectivity through  Landauer's formula \cite{landauer}:
\begin{eqnarray}
\rho\propto\frac{R}{T},
\label{resis}
\end{eqnarray}    
which is of great importance in  condensate matter and  it can be calculated
from a single particle theory with no dissipation \cite{das-green}.
From Eqs. (\ref{eq10a}) and (\ref{resis}) we conclude that the resistivity, for 
the  narrow channel, is:
\begin{equation}
\rho\propto G(r)a^{\frac{3}{2}}.
\end{equation}

Let us now briefly discuss  the case of the wide channel with small amplitudes. 
According to Eq. \cite{luna1} 
chaotic scattering on surfaces with deterministic profiles is practically indistinguishable 
from the scattering on surfaces with  random profiles. It is known that 
the classical resistivity $\rho$ of a plate with rough surfaces grows quadratically \cite{nandini,{meyerovich}}
with the root mean square (rms) height of the roughness $\xi$, when $\xi<<2b$.  By using
Landauer's formula,
we can express the transmitivity as:
\begin{equation}
T(\xi)=1-c\xi ^2,
\end{equation}
where $c$ is a constant  that depends on the geometrical properties of  the  channel. 
Assuming that the rms height of the effective random profile is 
proportional to the amplitudes of the ripples $a$, we conclude that  
\begin{equation}
R= F(r,L)a^2,
\label{random}
\end{equation}
and hence $\rho\propto F(r,L)a^2$. In contrast to the narrow channel, the function $F(r,L)$ remains unknown.

\section{Numerical results for the reflectivity and transmitivity}\label{numerical}

To obtain numerical results we considered  $N=10^2$ sources at $x_0=0$ and $n_0=10^5$,
with the  distribution given by  Eq. (\ref{distri1}). Our numerical method   
takes into account the
possibility of multiple collisions, by generalizing the map in Eq. (\ref{map})

\begin{figure}
 \centering
\subfigure[]{\includegraphics[scale=0.3]{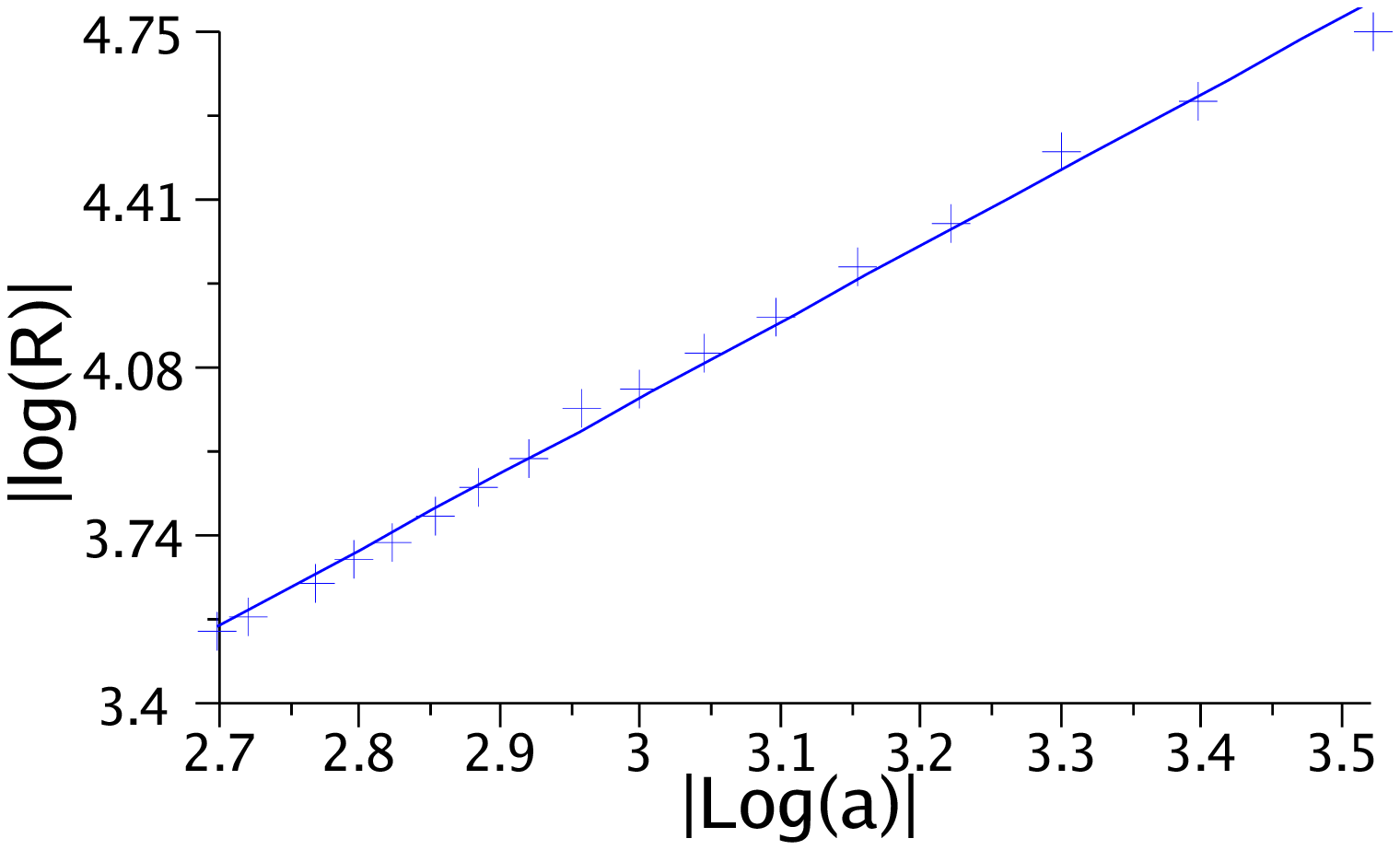}}\qquad
\subfigure[]{\includegraphics[scale=0.3]{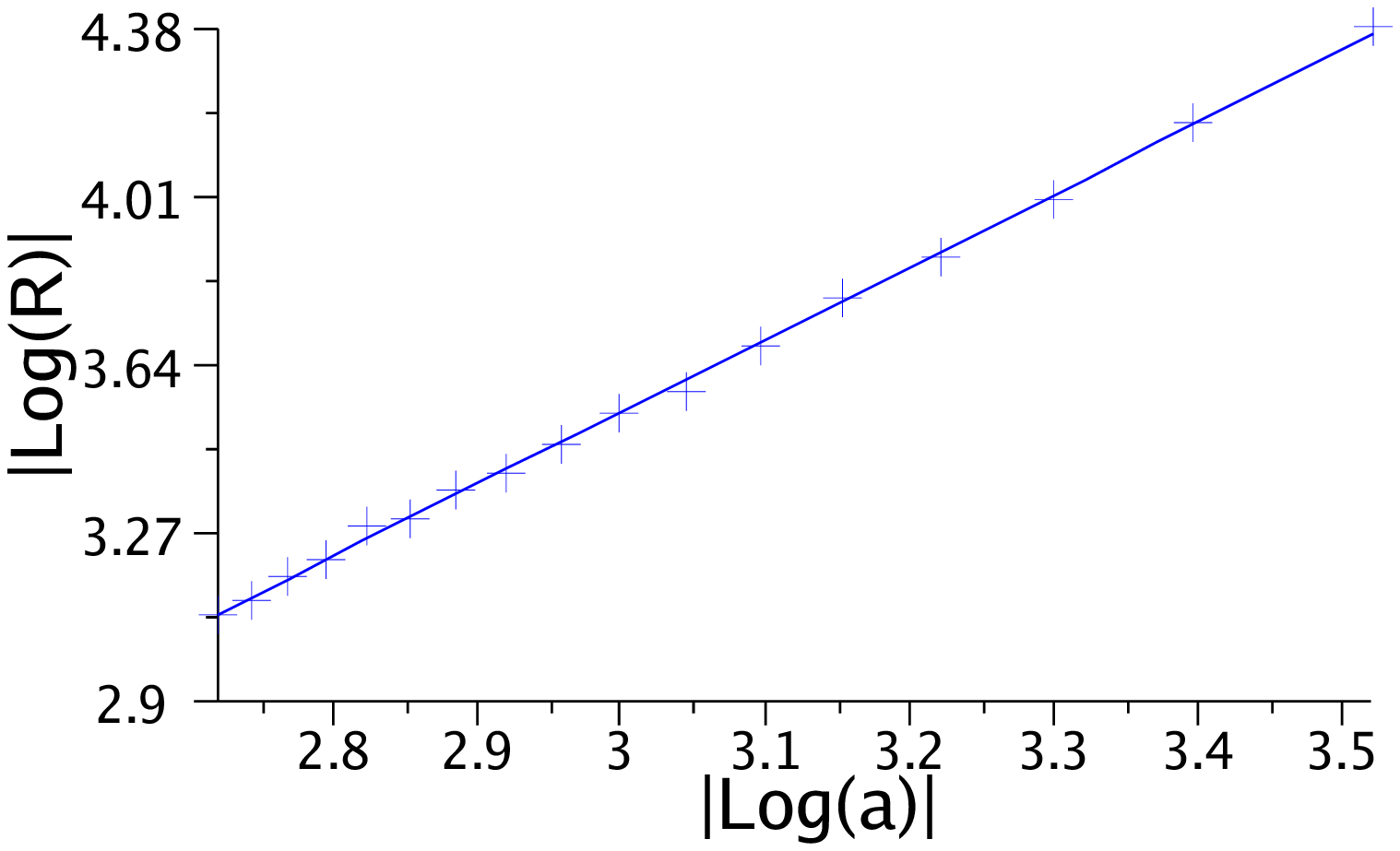}}\\
\subfigure[]{\includegraphics[scale=0.3]{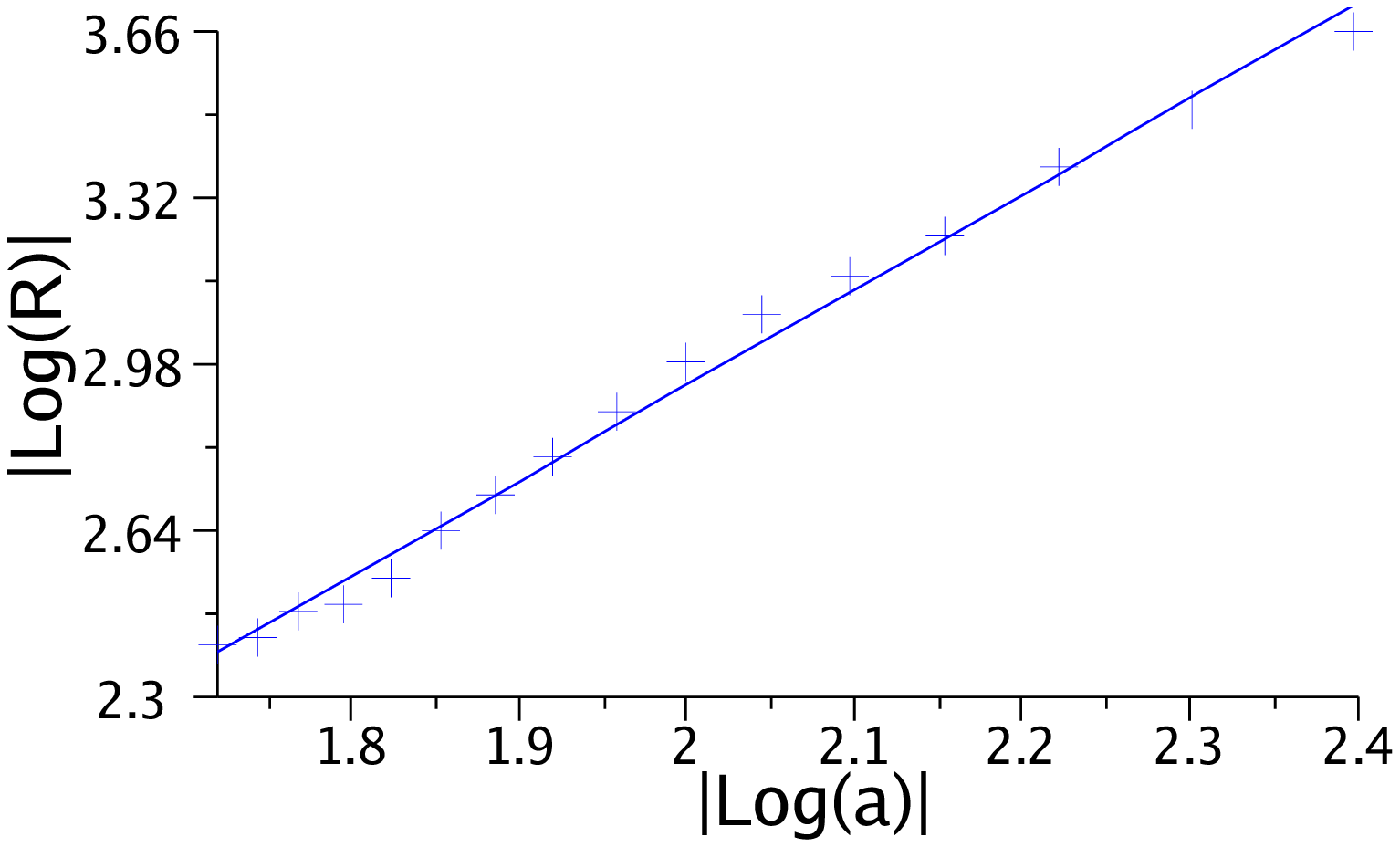}}\qquad
\subfigure[]{\includegraphics[scale=0.3]{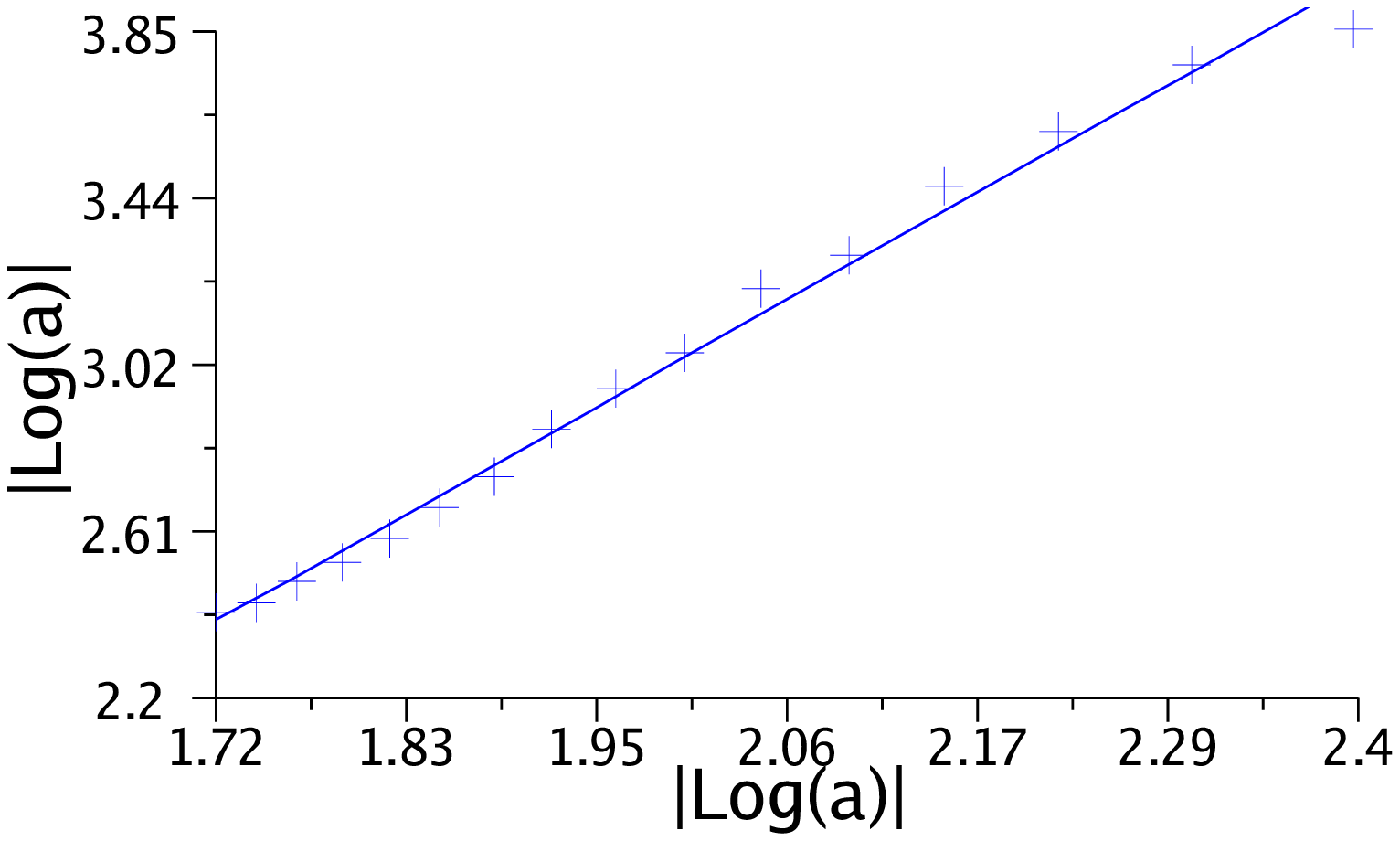}}\\
\caption{ 
Numerical results  of $|\log R|$ vs.   $|\log a|$  for the narrow 
channel (Figs. (a) and (b)).
 and  for the wide channel (Figs. (c) and (d)). In (a)  $r=\frac{1}{4}$, in  (b)   $r=\frac{1}{3}$,  in (c)  $r=\frac{1}{2}$,  and finally in (d)  $r=0$. }
\label{resis1}
\end{figure}

\begin{figure}
\centering
\includegraphics[scale=0.3]{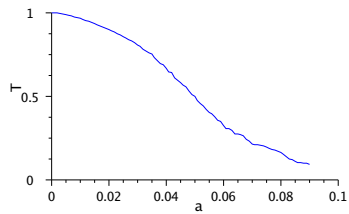}
\caption{ Numerical results for the transmitivity in the narrow channel  for $r=1/2$. }
\label{trans1}
\end{figure}

In order to check the confidence of our numerical method, first in Fig. \ref{resis1} we show the numerical results ($|\log R|$ vs.   $|\log a|$) with  cross symbols and  the corresponding linear  fit in solid line   for different values of $r$ and small amplitudes. 
In  Figs. \ref{resis1} (a) and (b), for the narrow channel, the fitting gives the results
$R\sim a^{1.59}$ for $r=\frac{1}{4}$ and $R\sim a^{1.58}$ for 
$r=\frac{1}{3}$. While for the wide channel we have $R\sim a^{1.9}$ for $r=\frac{1}{2}$ 
and $R\sim a^{2.15}$ for $r=0$. The corresponding theoretical results are $R\sim a^{1.5}$ for 
Figs. \ref{resis1} (a) and (b), and $R\sim a^2$ for Figs. \ref{resis1} (c) and (d).
According to  these  results    the agreement with the analytical  
 results in  Eqs. (\ref{eq10a}) and (\ref{random}) is good.
On the other hand, we calculate the  error,  $E=\sqrt{\sum_i(R_i^{th}-R_i^{num})^2}$, 
where $R_i^{th}$ are determined from Eq. (\ref{eq10a}) and $R_i^{num}$ are the numerical results using in Figs. \ref{resis1} (a) and (b). We obtain that $E=6.8\times 10^{-5}$
for $r=1/4$ and $E=1.5\times 10^{-4}$ for $r=1/3$, which are small.
Second, in Fig. \ref{trans1} we show the transmitivity for a narrow channel for $r=1/2$. We compare this
result with the Fig. 4 (c) presented in Ref. \cite{luna1} and observe a good agreement.

It is  known that   a cavity based on the narrow channel  has potential applications to chaotic waveguides and microlasers \cite{luna5}. For these applications it is important 
to have a big number of trapped particles (rays) which  give a high reflectivity.  Our channel  
is an example of this kind of systems.  As  already mentioned we introduced a phase 
shift, $r$,  as a new parameter which makes  the dynamics of these kind of channels more interesting. 
To see this we varied $r$ and observed  a considerable  increment of the
resistivity  comparing with   the semiplane channel, which is equivalent to our channel whenever  $r=1/2$. 

\begin{figure}[ht]
\begin{center}
\includegraphics[scale=0.3]{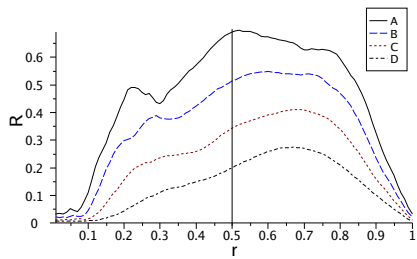}
\caption{\label{resisr} Reflectivity of the narrow channel vs phase for (A) $a=0.06$; 
(B)$a=0.05$; (C)$a=0.04$; (D)$a=0.03$} 
\end{center}
\end{figure}

In Fig. \ref{resisr} 
we show a plot for the reflectivity as a function of the phase shift, for different 
amplitudes of the ripple. In this figure we indicate reference points at each curve located at $r=0.5$ which
define the  $R_{ref}$. At these points, we observe that the resistivity 
is maximal for values of $r\neq 0.5$, consequently the resistivity of the channel
can be manipulated by varying the phase shift. If we define $R_{max}$ as the maximum
value of the reflectivity for a given $a$, we can introduce a parameter, $I$, that measures
the  relative increment of the  reflectivity.
\begin{equation}
I=\left|\frac{R_{max}-R_{ref}}{R_{ref}}\right|,
\end{equation}
The values of $I$  for the curves $A$, $B$, $C$ and $D$ are $0.009$, $0.087$, $0.230$, 
and $0.400$ respectively, which represent  increase of $0.9\%$,  $8.7\%$,  $23\%$ and  $40\%$.
Let us observe that bigger values  of $I$ result from smaller values  of the
amplitude (see curves $C$ and $D$ in Fig.\ref{resisr}). This result can be explained by 
observing that the rate of destruction of the librational orbits, as $a$ is increased, is slower for $r=1/2$ than for other cases. For the curves $C$ and $D$ in Fig.\ref{resisr}
the main contribution to the
reflectivity comes from  particles executing librational motion. To illustrate 
this  see Fig.\ref{lnc2} (c), from which we can see large regular regions corresponding 
to this kind of particles.

\section{Concluding remarks}\label{concluding}

We have studied  some of the transport properties of classical particles through a two-dimensional channel with  sinusoidal  boundaries in the ballistic regime. We restrict ourselves to cases 
of   narrow channels and   wide channels. In the first one, Poincar\'e plots show a 
regular dynamics for small amplitudes. A transition to mixed chaos is 
observed as the amplitude is increased. In the second case, chaotic behaviour 
appears even for small amplitudes. For bigger amplitudes global 
chaos is reached. In both cases the rate of the transition to chaos 
depended on the phase shift. For the narrow channel, the contributions of regular and chaotic regions 
to reflection were identified via Poincar\'e plots. 

An analytical  approximate expression for the classical resistivity  taking into account 
small ripple amplitudes was obtained. We found, 
that the resistivity $\rho$ behave like $\rho(a,r)\propto G(r)a^{\frac{3}{2}}$ in the case of narrow channel (showing  a regular dynamics) and for wide channel (showing a chaotic dynamics) 
$\rho(a,r,L)\propto F(r,L)a^2$, where $G(r)$ being a known function while 
$F(r,L)$ remains unknown. These results were corroborated by numerical calculations.

Taking as  parameters the amplitude 
of the ripples and the  phase shift between the boundaries, we observed that the manipulation 
of the phase shift results in a considerable  increment of the resistivity  when  
it is compared
with that obtained in the semiplane channel. This shows that the use of the phase shift 
allows favorable conditions  for potential applications in waveguides and resonators. 
For this applications  a complementary quantum  analysis is  necessary.

\section*{Acknowledgments}
 Authors  thank to CIC-UMSNH and COECYT for partial support.

\appendix 
\section{}\label{appen}

\begin{figure}[ht]
\begin{center}
\includegraphics[scale=0.3]{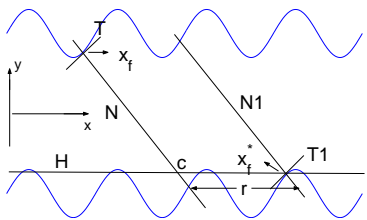}
\caption{\label{enfase} The channel for $r=0$.} 
\end{center}
\end{figure}

In order to find the localization of the fixed points we refer to  Fig. \ref{enfase} 
taking the case in which the upper and lower walls are in phase $r=0$, 
see Eq. (\ref{walls}).  The  fixed points  represent particles bouncing between  
the same points   with the upper and lower rippled boundaries respectively: $(x_f,y_1(x_f))$ and $(x_f^*,y_2(x_f^*))$. In order to find these points we must first find the points on the walls fulfilling the condition $\frac{dy_1}{dx}\mid_{x_f}=\frac{dy_2}{dx}\mid_{x_f^*,r=0}$, 
this leads to:
\begin{eqnarray}
\cos(2\pi x_f)&=&\cos(2\pi (x_f+h)),
\label{fijo6}
\end{eqnarray}
 where for convenience we write $x_f^*=x_f+h$. 

Let us consider two parallel lines $N$ and $N_1$. These lines intercept the curves $y_1$ 
and $y_2$ at the points $(x_f,y_1(x_f))$ and $(x_f^*,y_2(x_f^*))$. They are normal to the 
upper wall and lower wall at the points to the tangents $T$ and $T_1$ at $x_f$ and 
$x_f^*$ respectively. If $x_f$ is the $x$ position of a fixed point then $N$ and $N_1$ 
must be equal once we have made an horizontal shift between the walls by a quantity $r$ 
in our initial configuration (see Fig. \ref{enfase}). To find $r$, we must 
obtain the $x$ coordinate of the point $c$  ($x_c$) that is defined by the intercept between the lines $N$ and 
$H=y_2\mid_{x_f^*}$. Consequently $r$ is determined by: 
\begin{equation}
r=x_f^*-x_c.
\end{equation}
This leads to the condition for $r$
\begin{equation}
r=h-4\pi ba\cos(2\pi x_f)+2\pi a^2\cos2\pi x_f(\,{\rm sin}\,2\pi(x_f+h)-\,{\rm sin}\,2\pi x_f),
\label{fijo4}
\end{equation}
where $h$ is given by Eq. (\ref{fijo6}).
Finally the $x$ component of the momentum for fixed points is given by:
\begin{equation}
p_f=\cos\left( \tan^{-1}\left( -\frac{1}{\frac{dy_1}{dx}|_{x_f}}\right)\right).
\end{equation}
We can obtain the same result as the previously obtained through the method of the effective 
potential given in  Eq. (\ref{fijopeque}) by considering small amplitudes of the rippled  
in Eqs. (\ref{fijo6}) and (\ref{fijo4}).

\end{document}